\documentclass[12pt]{iopart}

  \usepackage{bm,amssymb}
  \usepackage{graphicx}
  \usepackage{cite}
  \usepackage{slashed}

\begin{document}

\title{$R_{\rm pA}$ ratio: total shadowing due to running coupling}

\author{E.~Iancu$^1$
and D.N.~Triantafyllopoulos$^2$}

\address{$^1$ Service de Physique Th\'eorique de Saclay,
F-91191 Gif-sur-Yvette, France}

\address{$^2$ ECT$^*$, Villa Tambosi, Strada delle Tabarelle 286,
I-38050 Villazzano (TN), Italy}

\begin{abstract}
We predict that the $R_{\rm pA}$ ratio at the most forward rapidities to
be measured at LHC should be strongly suppressed, close to ``total
shadowing'' ($R_{\rm pA}\simeq A^{-1/3}$), as a consequence of running
coupling effects in the nonlinear QCD evolution.

\end{abstract}


\noindent We present predictions for the nuclear modification factor, or
``$R_{\rm pA}$ ratio'', at forward pseudorapidities ($\eta>0$) and
relatively large transverse momenta ($p_{\perp}$) for the produced
particles, in the kinematical range to be accessible at LHC. These
predictions are based on a previous, systematic, study of the $R_{\rm
pA}$ ratio within the Color Glass Condensate formalism with running
coupling \cite{Iancu:2004bx}. The ratio can be approximated by
 \begin{eqnarray}\label{rpadef}
  R_{\rm pA} \simeq \frac{1}{A^{1/3}} \frac{\Phi_{\rm A}(Y,p_{\perp})}{\Phi_{\rm p}(Y,p_{\perp})},
 \end{eqnarray}
where $Y=\eta + \ln\sqrt{s}/p_{\perp}$ and $\Phi(Y,p_{\perp})$ is the
unintegrated gluon distribution of the corresponding target hadron at
fixed impact parameter. When the energy increases one expects more and
more momentum modes of this distribution to saturate to a value of order
$1/\alpha_s$, and the corresponding saturation momentum reads
 \begin{eqnarray}\label{Qsat}
   Q_s^2(Y) = \Lambda^2
   \exp{\sqrt{B (Y-Y_0) + \ln^2\frac{Q_s^2(Y_0)}{\Lambda^2}}},
 \end{eqnarray}
with $\Lambda=0.2 {\rm GeV}$, $B=2.25$ and $Y_0=4$. The initial condition
for the nucleus and the proton are taken as $Q_s^2(A,Y_0) = 1.5 {\rm
GeV}^2$ and $Q_s^2(p,Y_0) = 0.25 {\rm GeV}^2$ respectively, so that
$Q_s^2(A,Y_0) = A^{1/3} Q_s^2(p,Y_0)$ for $A=208$. The functional form of
this expression is motivated by the solution to the nonlinear QCD
evolution equations with running coupling
\cite{Mueller:2002zm,Mueller:2003bz}, while the actual values of the
numbers $B$ and $Y_0$ have been chosen in such a way to agree with the
HERA/RHIC phenomenology. As shown in Fig.~1, with increasing $Y$ the two
saturation momenta approach to each other and clearly for sufficiently
large $Y$, a nucleus will not be more dense than a proton
\cite{Mueller:2003bz}.

\begin{figure}[ht]\label{Fig1}
\centerline{\includegraphics[width=7.5cm]{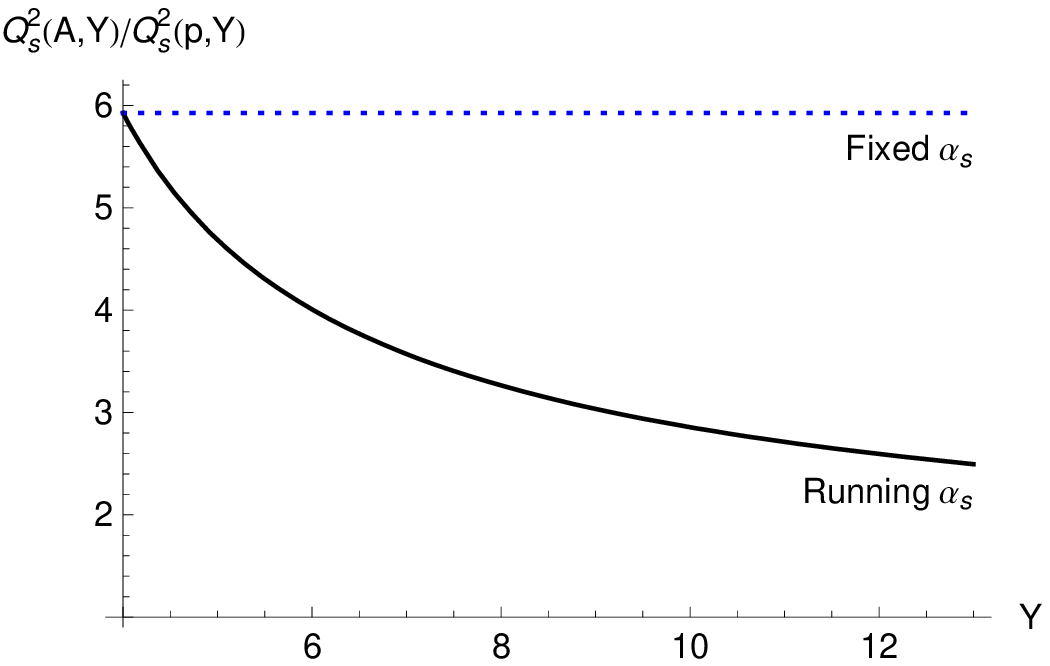}\hspace{0.5cm}
    \includegraphics[width=7.5cm]{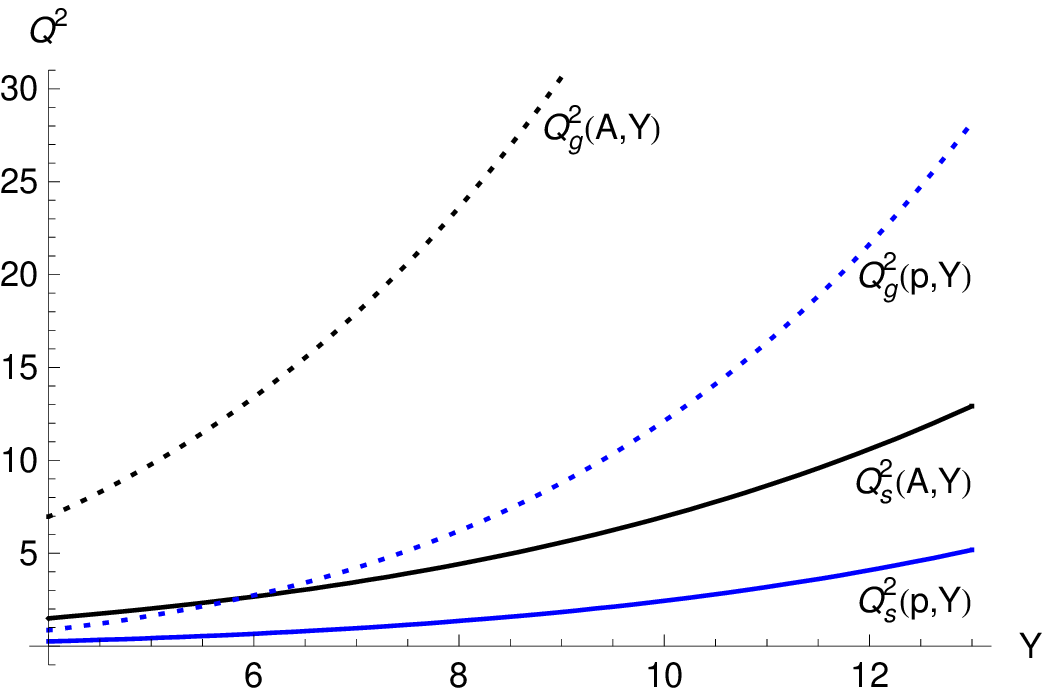}}
    \caption{Left: The ratio of the saturation momenta.
    ($Y=12$ corresponds to a pseudorapidity
    $\eta=6$ for the produced particles). Right: Geometric scaling windows.}
\vspace{-0.5cm}
\end{figure}

For momenta $p_{\perp}$ larger than $Q_s$, the gluon distribution
satisfies geometrical scaling \cite{Iancu:2002tr,Mueller:2002zm}, i.e.~it
is a function of only the combined variable $p_{\perp}/Q_s(Y)$ :
  \begin{eqnarray}\label{phi}
   \Phi(p_{\bot},Y) \propto \left[\frac{Q_s^2(Y)}{p_{\bot}^2}
   \right]^{\gamma} \left( \ln\frac{p_{\bot}^2}{Q_s^2(Y)} + c \right),
 \end{eqnarray}
with $\gamma=0.63$ and $c=\mathcal{O}(1)$. This holds within the scaling
window $Q_s \lesssim p_{\perp} \lesssim Q_g$, where $\ln
Q_g^2(Y)/Q_s^2(Y) \sim [\ln Q_s^2(Y)/\Lambda^2]^{1/3}$ and for large $Y$
this is proportional to $Y^{1/6}$. The geometrical scaling lines for a
proton and a nucleus are shown in Fig.~1. Note that, since $Q_g$ is
increasing much faster than $Q_s$, a {\em common scaling window} exists,
at $Q_s(A,Y) \lesssim p_{\perp} \lesssim Q_g(p,Y)$ (and for sufficiently
large $Y$), where the gluon distributions for both the nucleus and the
proton are described by Eq.~(\ref{phi}).

Within this window, it is straightforward to calculate the $R_{\rm pA}$
ratio. This is shown in Fig.~2 for two values of pseudorapidity. The
upper, dotted, line is the asymptotic prediction of a fixed-coupling
scenario, in which the ratio ${Q_s^2(A,Y)}/{Q_s^2(p,Y)}={\rm const.}=
A^{1/3}$, while the lowest, straight, curve is the line of total
shadowing $R_{\rm pA}=1/A^{1/3}$. Our prediction with running coupling is
the line in between and it is very close to total shadowing. This is
clearly a consequence of the fact that the proton and the nuclear
saturation momenta approach each other with increasing energy.

\begin{figure}[ht]\label{Fig2}
    \centerline{\includegraphics[width=7.5cm]{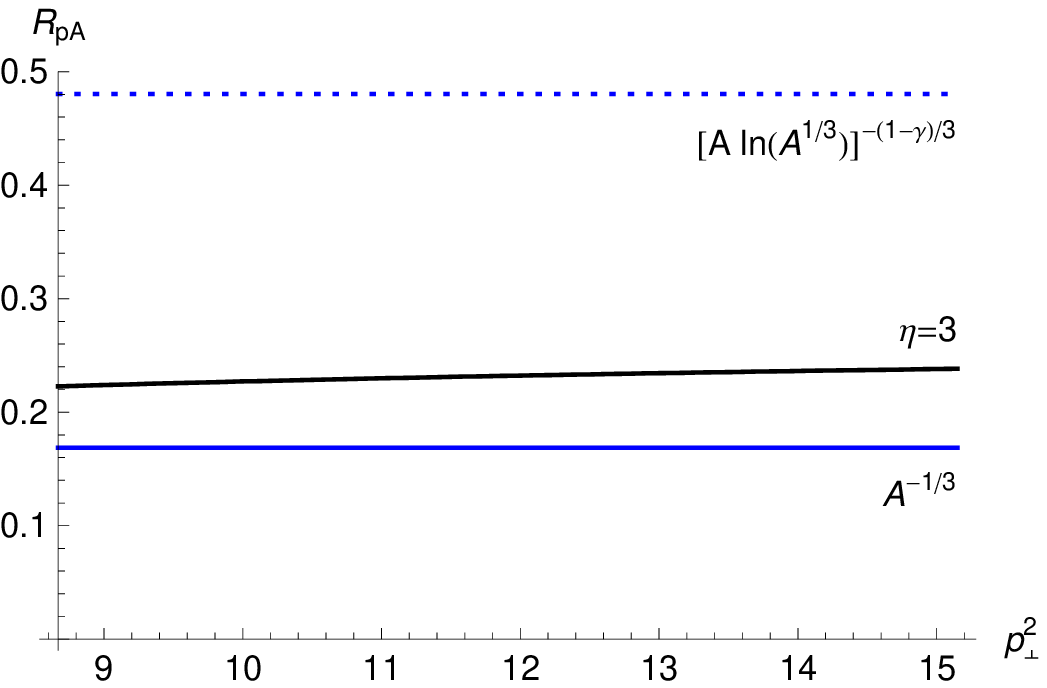}\hspace{0.5cm}
    \includegraphics[width=7.5cm]{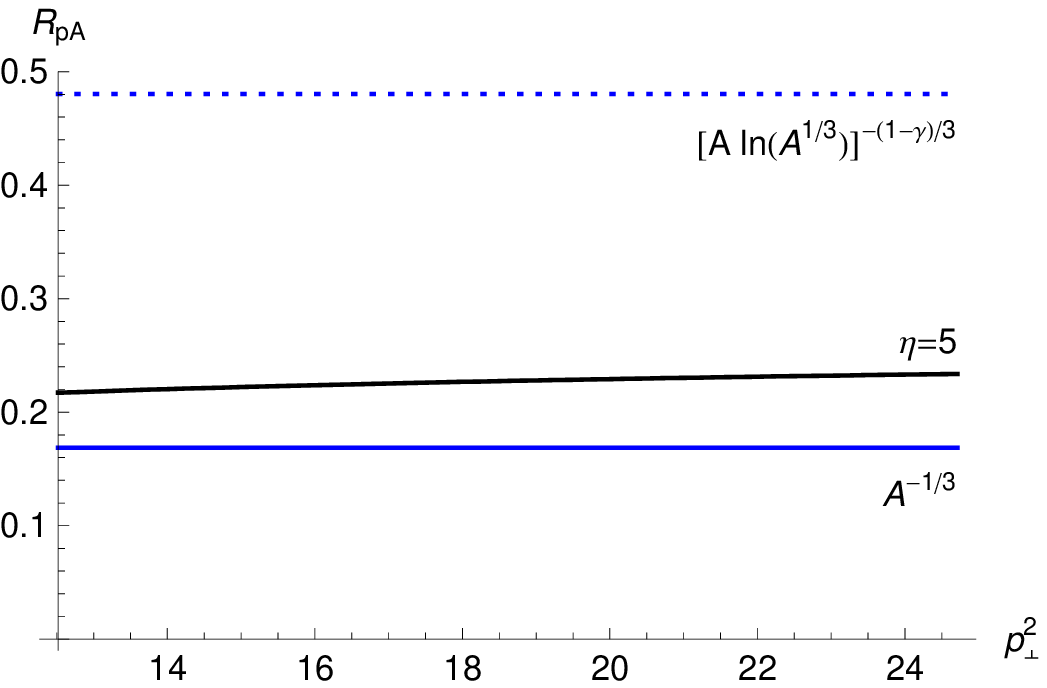}}
    \caption{The ratio $R_{\rm pA}$ as a function of $p_{\perp}^2$ at $\sqrt{s}=8.8\,$TeV.}
\vspace{-0.4cm}
\end{figure}

Note finally that in the present analysis we have neglected the effects
of particle number fluctuations (or ``Pomeron loops''). This is
appropriate since Pomeron loops effects are suppressed by the running of
the coupling \cite{Dumitru:2007ew}, and thus can be indeed ignored at all
energies of phenomenological interest (in particular, at LHC).

\end{document}